\documentclass[aps,preprint,prd,showpacs,nofootinbib]{revtex4}
\usepackage{amsmath}
\usepackage{graphicx}
\usepackage{subfigure}
\usepackage{multirow}
\usepackage{bm}
\usepackage{amssymb}
\usepackage{mathtools}
\usepackage{enumerate}
\usepackage{ulem}
\usepackage{color}
\usepackage[colorlinks,linkcolor=magenta,anchorcolor=cyan,citecolor=blue]{hyperref}

\def\be{\begin{equation}}
\def\ee{\end{equation}}

\begin{document}
\title{Towards inflation with $n_s=1$ in light of Hubble tension and primordial gravitational waves}

\author{Gen Ye$^{1,2}$\footnote{yegen14@mails.ucas.ac.cn}}
\author{Jun-Qian Jiang$^{2}$\footnote{jqjiang@zju.edu.cn}}
\author{Yun-Song Piao$^{1,2,3,4}$\footnote{yspiao@ucas.ac.cn}}

\affiliation{$^1$ School of Fundamental Physics and Mathematical
    Sciences, Hangzhou Institute for Advanced Study, UCAS, Hangzhou
    310024, China}

\affiliation{$^2$ School of Physics, University of Chinese Academy
of Sciences, Beijing 100049, China}

\affiliation{$^3$ International Center for Theoretical Physics
    Asia-Pacific, Beijing/Hangzhou, China}

\affiliation{$^4$ Institute of Theoretical Physics, Chinese
    Academy of Sciences, P.O. Box 2735, Beijing 100190, China}

\begin{abstract}

Recently, it has been found that complete resolution of the Hubble
tension might point to a scale-invariant
Harrison-Zeldovich spectrum of primordial scalar perturbation,
i.e. $n_s=1$ for $H_0\sim 73$km/s/Mpc. We show that for well-known
slow-roll models, if inflation ends by a waterfall instability
with respect to another field in the
field space while inflaton is still at a deep slow-roll region,
$n_s$ can be lifted to $n_s= 1$. A surprise of our
result is that with pre-recombination early dark energy, chaotic
$\phi^2$ inflation, ruled out by Planck+BICEP/Keck in
standard $\Lambda$CDM, can be revived, which is now well within
testable region of upcoming cosmic microwave background B-mode
experiments.

\end{abstract}

\maketitle

\section{Introduction}

Inflation
\cite{Guth:1980zm,Linde:1981mu,Albrecht:1982wi,Starobinsky:1980te,Linde:1983gd}
is the current paradigm of early universe, which predicts (nearly)
scale-invariant scalar perturbation, as well as primordial
gravitational waves (GW). In well-known single field slow-roll
inflation models, the spectral index $n_s$ of primordial scalar
perturbation follows
\cite{Mukhanov:2013tua,Kallosh:2013hoa,Roest:2013fha,Martin:2013tda}
\be n_s-1 = -{{\cal O}(1)\over N_*} \label{ns}\ee in large $N_*$
limit, where $N_*=\int Hdt$ is the efolds number before the end of
inflation. The cosmic microwave background (CMB) perturbation
modes exit the horizon at about $N_*\sim60$ efolds, if
inflation ends around $\sim10^{15}\text{GeV}$. Recently, based on
standard $\Lambda$CDM model the Planck collaboration obtains $n_s\approx 0.97$ \cite{Planck:2018vyg}, which is
consistent with (\ref{ns}).

However, the current expansion rate of the Universe, Hubble
constant $H_0$, inferred by the Planck collaboration
\cite{Planck:2018vyg} assuming $\Lambda$CDM is in $\gtrsim
5\sigma$ tension with that reported recently by the SH0ES
collaboration \cite{Riess:2021jrx} using Cepheid-calibrated
supernovas. Currently, it is arriving at a consensus that this
so-called Hubble tension likely signals new physics beyond $\Lambda$CDM
\cite{Verde:2019ivm,Knox:2019rjx}, see also
Refs.\cite{DiValentino:2021izs,Perivolaropoulos:2021jda, Dainotti:2021pqg, Dainotti:2022bzg} for
reviews and some recent developments.

As a promising resolution of Hubble tension, early
dark energy (EDE) \cite{Karwal:2016vyq,Poulin:2018cxd}, which is
non-negligible only for a few decades before
recombination\footnote{Actually, ``EDE" corresponds to
EDE+$\Lambda$CDM, which is a pre-recombination modification to
$\Lambda$CDM model, and the evolution after recombination must
still be $\Lambda$CDM-like.}, has been extensively studied
e.g.\cite{Agrawal:2019lmo,Lin:2019qug,Smith:2019ihp,Niedermann:2019olb,Sakstein:2019fmf,Ye:2020btb,Ye:2020oix,Lin:2020jcb,Seto:2021xua,Karwal:2021vpk,Vagnozzi:2021gjh}.
In original (axion-like) EDE \cite{Poulin:2018cxd}, the scalar
field with $V(\varphi)\sim \left(1-\cos(\varphi/f_a)\right)^3$ is
responsible for EDE, which starts to oscillate at critical
resdshift, and dilutes away rapidly like a fluid with $w>1/3$
before recombination. In AdS-EDE \cite{Ye:2020btb}, since the
potential has an anti-de Sitter (AdS) well
(temporarily realizing $w>1$ so EDE dilutes away
faster), a larger EDE fraction and so higher $H_0(\approx
73$km/s/Mpc) can be achieved without spoiling fit to
fullPlanck+BAO+Pantheon dataset. Recently, combined analysis of
Planck ($\ell_\mathrm{TT} \lesssim 1000$) with ACT and SPT data
for EDE has also been performed, such as Planck+SPTpol
\cite{Chudaykin:2020acu,Chudaykin:2020igl,Jiang:2021bab},
Planck+ACT DR4 \cite{Hill:2021yec,Poulin:2021bjr} and Planck+ACT
DR4+SPT-3G \cite{LaPosta:2021pgm,Smith:2022hwi,Jiang:2022uyg},
see also
\cite{Hill:2020osr,Ivanov:2020ril,DAmico:2020ods,Ye:2021iwa} for
Planck+large scale structure data.

\begin{figure}[!h]
\includegraphics[width=0.8\linewidth]{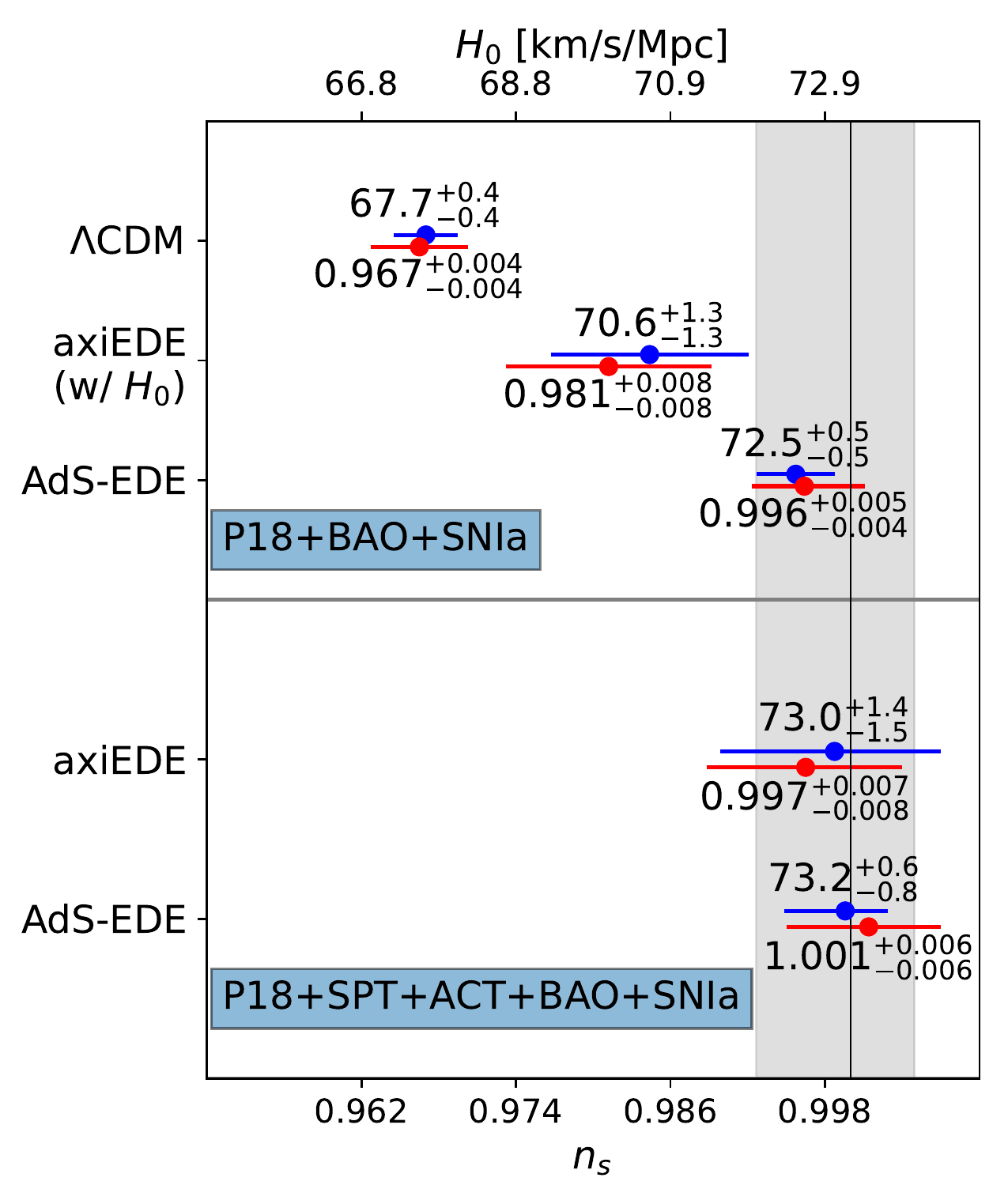}
\caption{ $n_s$ vs. $H_0$. In upper panel with the
fullPlanck+BAO+Pantheon dataset, we adopt the result in
Ref.\cite{Poulin:2018cxd} for original axion-like EDE (the SH0ES
result as a Gaussian prior on $H_0$), and Ref.\cite{Jiang:2021bab} for
AdS-EDE. In lower panel with the Planck+ACT+SPT+BAO+Pantheon
dataset, Ref.\cite{Smith:2022hwi} (Planck $\ell_\mathrm{TT}<650$)
for axion-like EDE and Ref.\cite{Jiang:2022uyg} (Planck
$\ell_\mathrm{TT}<1000$) for AdS-EDE. Grey band represents the recent SH0ES
result $H_0= 73.04\pm1.04$km/s/Mpc \cite{Riess:2021jrx}, and black
solid line marks $n_s=1$. }
    \label{fig:h0ns}
\end{figure}

In Ref.\cite{Ye:2021nej}, it has been found that in corresponding
Hubble-tension-free cosmologies, the bestfit values of
cosmological parameters acquired assuming $\Lambda$CDM must shift
with $\delta H_0$, and with fullPlanck+BAO+Pantheon dataset the
shift of $n_s$ scales as
 \be {\delta n_s}\simeq 0.4{\delta
H_0\over H_0}, \label{deltans}\ee
which suggests that
pre-recombination resolution of Hubble tension is pointing to a
scale-invariant Harrison-Zeldovich primordial spectrum, i.e.
$n_s=1$ for $H_0\sim73$km/s/Mpc, see also \cite{Benetti:2017gvm,Benetti:2017juy,Graef:2018fzu} for earlier discussion regarding $N_{eff}$ and $n_T$. The new constraint on tenor-to-scalar ratio $r$ with recent BICEP/Keck
data has also been recently considered in Ref.\cite{Ye:2022afu}. In Ref.\cite{Smith:2022hwi,Jiang:2022uyg}, with
Planck+ACT+SPT+BAO+Pantheon dataset, similar results have also
been found. We outlined the relevant results in
Fig.\ref{fig:h0ns}. Thus it is significant to explore the
implication of $n_s=1$ on primordial Universe.

How $n_s=1$ would affect our understanding about inflation? At
first thought, it seems that (\ref{ns}) is not compatible with the
result in Hubble-tension-free cosmologies, since for $N_*\approx
60$ it is hardly possible to achieve $n_s=1$ in the slow-roll
models satisfying (\ref{ns}). Actually, $n_s\geqslant 0.99$ puts a
lower bound $N_*> {\cal O}(10^2)$. Such perturbation
modes are still far larger than our observable Universe today. This seems to
pose a serious challenge to slow-roll inflation, implying that
corresponding models might need to be reconsidered
e.g.\cite{Barrow:1993zq,Vallinotto:2003vf,Starobinsky:2005ab}, see
also recent \cite{Takahashi:2021bti,DAmico:2021zdd,Lin:2022gbl}.

However, inspired by recent Ref.\cite{Kallosh:2022ggf}, we might
have a different story. In (\ref{ns}), $N_*$ is the ``distance"
between $\phi_*(\epsilon\ll 1)$ and $\phi_e(\epsilon=1)$ at which
inflation ends, $\phi$ being the inflaton. Typically $N_*\approx 60$. However, inflation can also be terminated at $\phi_c$ when $\epsilon\ll 1$ by waterfall instability with respect to another field $\sigma$, like in the
hybrid inflation models \cite{Linde:1991km,Linde:1993cn}, which suggests that $\Delta N\approx 60$ dose not necessarily require $N_*\approx 60$, see
Fig.\ref{fig:potential}. Thus we can actually
have $n_s$ arbitrarily close to $1$ by pushing $N_*$ to a
sufficiently large value $N_*\gg \Delta N\approx 60$ while ending
inflation by certain mechanism at $N_*-60$.

We will present this possibility. In our (hybrid) uplift of
$n_s\approx0.97$ to $n_s= 1$, the potential of inflaton $\phi$ still preserves the shape of well-known single field slow-roll inflation models, see
section-\ref{sec:model_data}, but inflation ends by the
waterfall instability while $\phi$ is still in the deep
slow-roll region. A surprise of our result is that
certain models originally thought to be ruled out by
Planck+BICEP/Keck based on $\Lambda$CDM \cite{BICEP:2021xfz},
specially chaotic $\phi^2$ inflation \cite{Linde:1983gd}, can be
revived by this hybrid uplift to $n_s= 1$, which is now well within the
testable region of upcoming CMB B-mode experiments, such as BICEP
Array \cite{Moncelsi:2020ppj} and CMB-S4 \cite{CMB-S4:2020lpa}.

\section{Hybrid uplift to $n_s=1$}\label{sec:model_data}

\begin{figure}
    \subfigure{\includegraphics[width=0.48\linewidth]{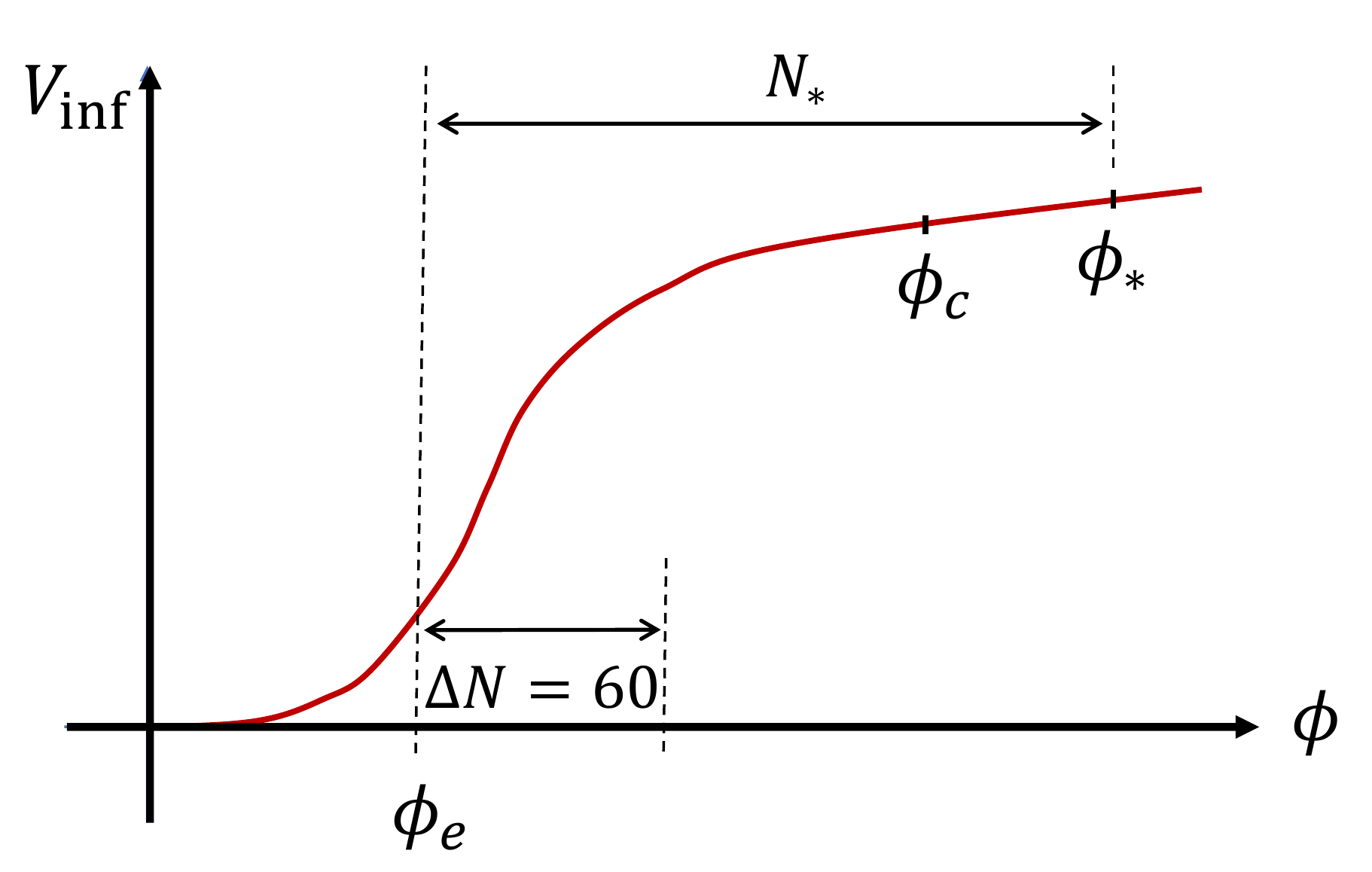}}
    \subfigure{\includegraphics[width=0.48\linewidth]{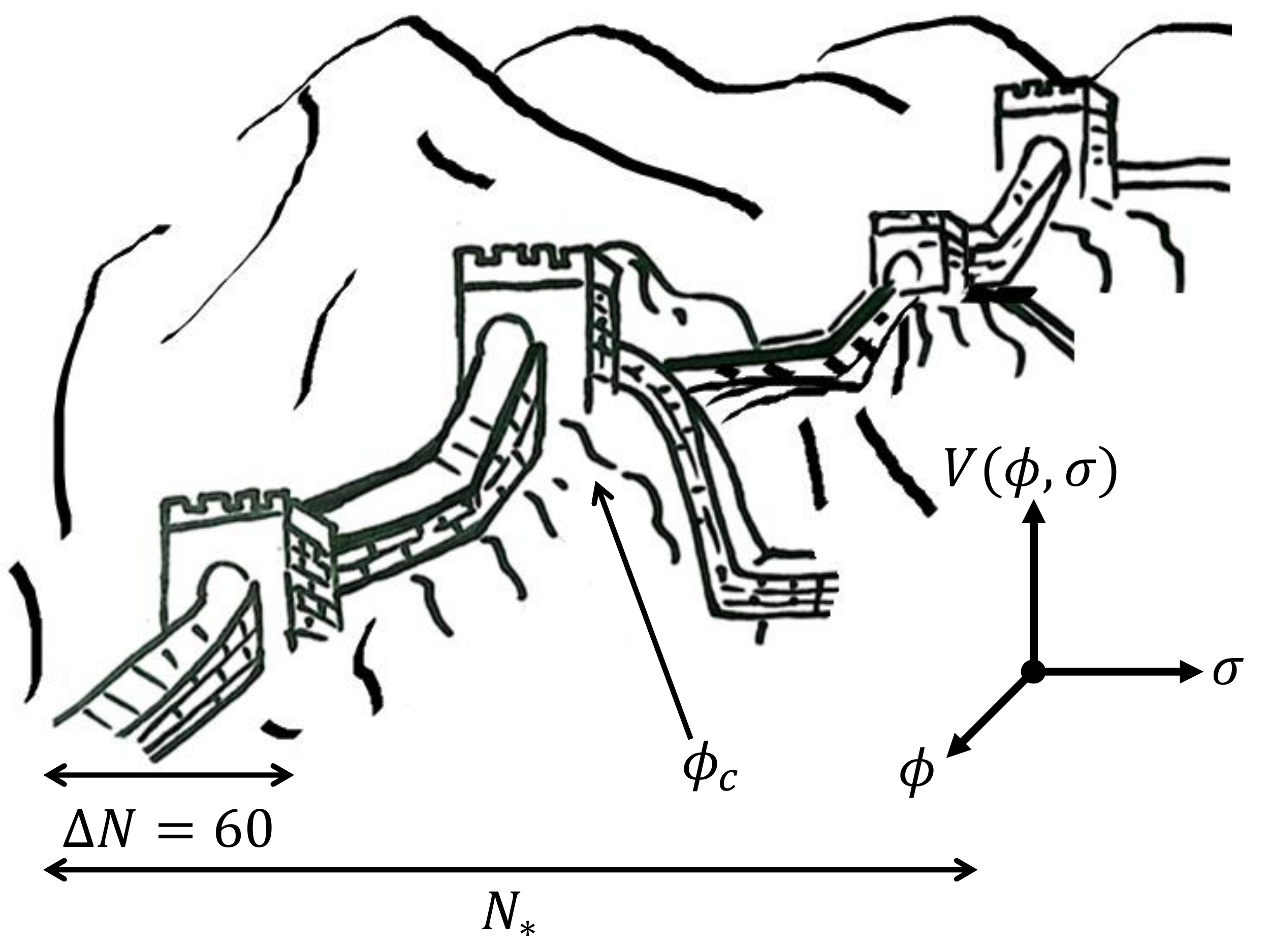}}
\caption{ \textit{Left panel}: The slow-roll potential $V_{\rm{inf}}$. \textit{Right panel}: $V(\phi,\sigma)$ obtained by hybrid lifting $V_{\rm{inf}}$ according to \eqref{V}. $\Delta N$ is the efolds number in the
original slow-roll model. In well-known slow-roll models,
inflation ends at $\phi_e$ where $\epsilon\simeq 1$. The
perturbation modes exiting horizon near $N_*\gg \Delta
N \approx 60$ is still far outside of our current Hubble horizon.
However, after the hybrid uplift of $V_{\rm{inf}}$ to the
$\phi-\sigma$ space (initially $\partial_\sigma^2V>0$ at
$\sigma=0$), inflation can end at $\phi=\phi_c$, at which we still
have $\epsilon(\phi_c)\ll 0.01$ but
$\partial_\sigma^2V<0$, by a waterfall instability along
$\sigma$, see the right panel, so that the modes
exiting horizon near $N_*\gg 60$ can be just at CMB
window, thus $n_s= 1$ in light of (\ref{ns}). }
    \label{fig:potential}
\end{figure}

The scenario we consider is sketched in Fig.\ref{fig:potential},
in which \be V(\phi,\sigma)= V_{\rm{inf}}(\phi)+{1\over
4\lambda}\left[\left(\lambda\sigma^2-M^2\right)^2-M^4\right]+{g^2\over
2}\sigma^2\phi^2, \label{V}\ee and $V_{\rm{inf}}$ is the
well-known inflation potentials satisfying (\ref{ns}). Initially,
$\sigma=0$ and $\partial_\sigma^2V>0$, the inflaton $\phi$ slow
rolls along $V_{\rm{inf}}$ and $\epsilon\ll 0.01$. At
$\phi=\phi_c\simeq {M/g}$, we still have $\epsilon(\phi_c)\ll
0.01$, but $\partial_\sigma^2V\lesssim 0$ so that the inflation
will rapidly end by a waterfall instability along $\sigma$
\footnote{At the minima of $\sigma$ when $\phi<\phi_c$,
the corresponding potential
$V=V_{\rm{inf}}-(M^2-g^2\phi^2)^2/4\lambda$ might be negative for
$|\phi|\ll M/g$ and certain $V_{\rm{inf}}$. In this case \eqref{V}
should be thought of as an effective potential only
captures the shape of field space for $\phi>\phi_c$ and $\phi\sim
\phi_c$.}, see also \cite{Romano:2018frb,Romano:2020oov} for the effective energy momentum tensor approach of multi-fields perturbations. In original hybrid inflation
\cite{Linde:1991km,Linde:1993cn}, when $\sigma=0$, $V=
V_{\rm{inf}}+{M^4\over 4\lambda}$ is lifted by $V_{\rm{up}}={M^4/
4\lambda}$. Generally, for $V_{\rm{inf}}\sim \phi^2$, when
$V_{\rm{inf}}\ll V_{\rm{up}}$, one has $n_s-1>0$
\cite{Copeland:1994vg}. However, here we subtract out the uplift
$V_{\rm{up}}$. We will see that for $V_{\rm{inf}}\sim \phi^2$, we
have $n_s-1\approx 0$ but $<0$.

In slow-roll approximation, one has ($M_p=1$) 
\be N_*\approx\Delta
N+\int_{\phi_e}^{\phi_c}{d\phi\over \sqrt{2\epsilon}}=
\left(\int_{\phi_c}^{\phi_*}+\int_{\phi_e}^{\phi_c}\right){d\phi\over
\sqrt{2\epsilon}}\approx N(\phi_*). \label{N}\ee
The
results of both $n_s$ and $r$ are determined by $\phi_*$, value of
the inflaton field $\phi$ when the corresponding perturbation mode
exits horizon during inflation, thus they are related to $N_*$
rather than $\Delta N$. This indicates that we can have $N_*>{\cal
O}(10^2)$ in (\ref{ns}) and $n_s\simeq1$, while still having
$\Delta N\approx 60$. In certain sense, with
(\ref{V}), what we do corresponds to push the inflaton $\phi$
deeply into slow-roll region at which $N_*\gg \Delta N\approx 60$.
Inflation will end at $N_*-60$ so that the modes exiting horizon
near $N_*$ can be just at CMB window.

Here, we show that the addition of a potential uplift
$V_{\rm{up}}$ in Ref.\cite{Kallosh:2022ggf} is actually equivalent
to pushing the inflaton $\phi$ deeply into the slow-roll region
without $V_{\rm{up}}$, i.e. eq.\eqref{V}. By lifting
$V_{\rm{inf}}=V_0\left(1-e^{-\gamma \phi}\right)$ to
$V_{\rm{up}}+V_{\rm{inf}}$, Ref.\cite{Kallosh:2022ggf} found
\begin{equation}\label{Linde}
    n_s-1\approx-{2\over \left({V_{\rm{up}}+V_{0}\over
\gamma^2V_0}\right)e^{\gamma\phi_*}}\approx-\frac{2}{\gamma^{-2}e^{\gamma\tilde{\phi}_*}}=-\frac{2}{\Delta
N+\gamma^{-2}e^{\gamma\tilde{\phi}_c}},\qquad
\tilde{\phi}\equiv\phi+{1\over \gamma}\ln(V_{\rm{up}}/V_0),
\end{equation}
where the second approximate equality is obtained in the large
$V_{\rm{up}}\gg V_0$ limit in the uplifted potential
$V_{\rm{up}}+V_{\rm{inf}}$. This is simply equivalent to the large
$N_*$ (or equivalently large $\phi_c$ and $\phi_*$) limit in
$V_{\rm{inf}}$ (without $V_{\rm{up}}$) because in light of
\eqref{N} we have
\begin{equation}\label{KL2}
    N_*\approx \int_{\phi_e}^{\phi_*}{d\phi\over
    \sqrt{2\epsilon}}= \gamma^{-2}e^{\gamma\phi}\Big|_{e}^{*},
\end{equation}
where $\epsilon={V_{\rm{inf}}^{\prime 2}\over 2V_{\rm{inf}}^2}
={\gamma^2\over 2}e^{-2\gamma\phi}$ is used. Thus combining
\eqref{KL2} with \eqref{Linde}, we obtain \eqref{ns}, which
indicates that large $V_{\rm{up}}$ or $\phi_c$ limit in
Ref.\cite{Kallosh:2022ggf} actually is equivalent to the large
$N_*$ limit. However, here we straightly push $N_*\approx 60$ to a sufficiently large value $N_*>{\cal
O}(10^2)$, and do not make the uplift of $V_{\rm{inf}}$ to
$V_{\rm{up}}+V_{\rm{inf}}$, so fully preserve the shape of single
field slow-roll potentials satisfying \eqref{ns}, thus \eqref{ns}
can be directly applied. The advantage of our inflation model will
be seen in the $\phi^p$ inflation.

\textbf{(hybrid) Starobinski inflation} In Starobinsky ($R^2$) model
\cite{Starobinsky:1980te}, the effective potential is
$V_{\rm{inf}}(\phi)\sim \left(1-e^{-\sqrt{2/3}\phi }\right)^2$, and
\begin{equation}
    n_s-1\approx -\frac{2}{N_*},\qquad r\approx\frac{12}{N^2_*},
\label{hybridS}\end{equation} which corresponds to $\alpha=1$ in
$\alpha$-attractor inflation
\cite{Kallosh:2013hoa,Ferrara:2013rsa,Kallosh:2013yoa,Galante:2014ifa}.
(\ref{hybridS}) is compatible with the $\Lambda$CDM constraint
$n_s\approx0.967$ for $N_*=\Delta N\approx 60$ in standard slow-roll inflation. However, in
Hubble-tension-free cosmologies, e.g.AdS-EDE, $n_s=0.998\pm0.005$,
see Fig.\ref{fig:h0ns} for the fullPlanck+BAO+Pantheon result,
this would require that in the hybrid lifted Starobinski model, we need $N_*\gtrsim 300\gg \Delta N$ and inflation
ending at $N\approx N_*-60$, thus $n_s\simeq1-2/N_*\gtrsim
0.993$ and $r\simeq{12\over N^2_*}\lesssim 1.3\times10^{-4}$. This tensor-to-scalar ratio is far
smaller than that in the standard Starobinsky model.

\textbf{(hybrid) $\phi^p$ inflation} In corresponding models,
$V_{\rm{inf}}(\phi)\sim \phi^p$, and
\begin{equation}
    n_s-1\approx -\frac{p/2+1}{N_*},\qquad r\approx\frac{4p}{N_*},
\end{equation}
Here, $p=2$ is the chaotic inflation \cite{Linde:1983gd},
$p=2/3,1$ correspond to the monodromy inflation
\cite{Silverstein:2008sg,McAllister:2008hb}. We have $n_s\approx
1-2/60=0.97$ for $p=2$ and $N_*=60$, which seems compatible with
the $\Lambda$CDM constraint. However, since $r\approx
8/{N_*}=0.13$, $\phi^2$ model has been ruled out by
Planck+BICEP/Keck data in $\Lambda$CDM \cite{BICEP:2021xfz}.

\begin{figure}
    \includegraphics[width=0.8\linewidth]{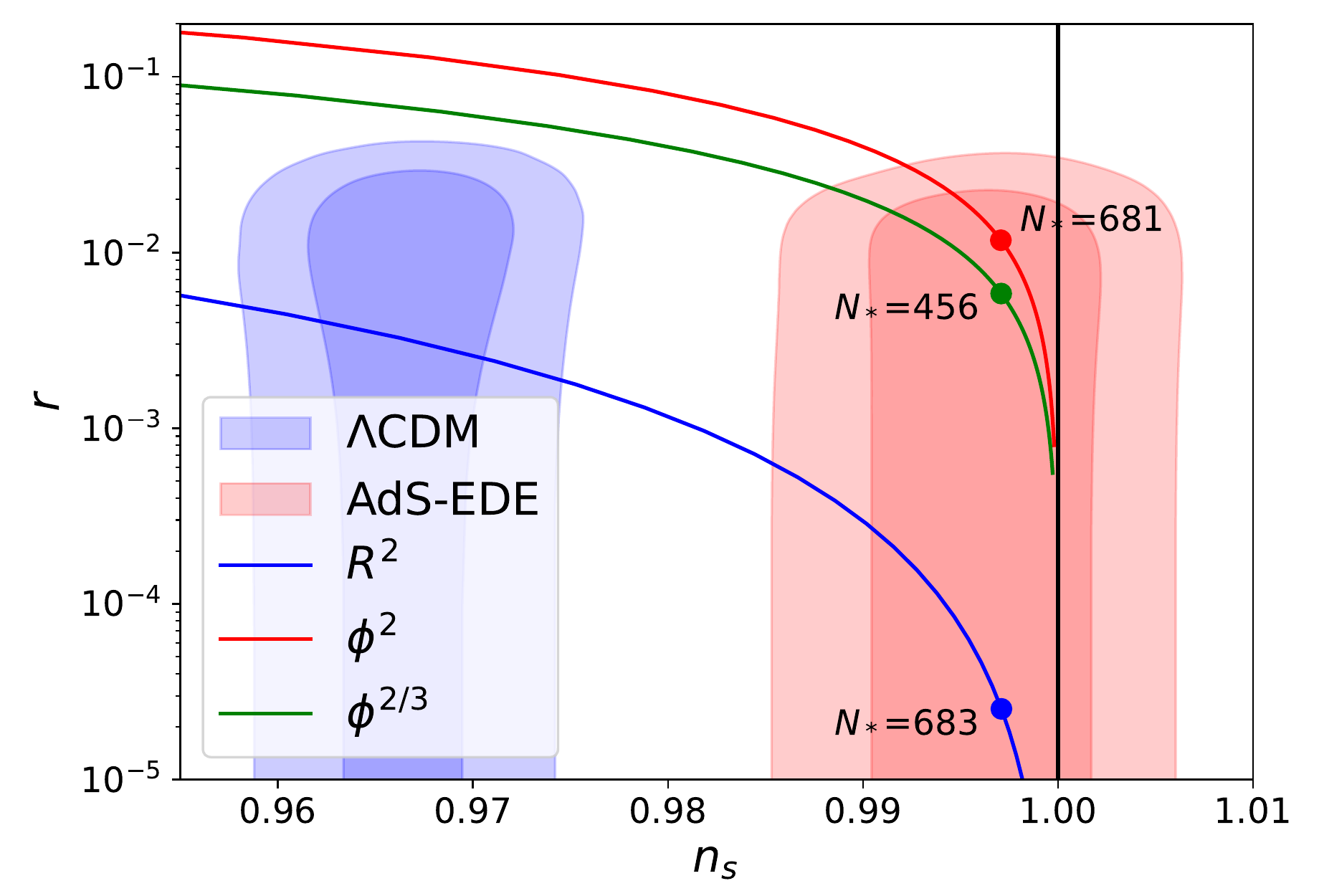}
\caption{Predictions of Starobinsky and $\phi^p$ ($p=2,2/3$)
models with respect to 68\% and 95\% C.L. contour of $r-n_s$.
Here, we adopt the result in recent BICEP/Keck
Ref.\cite{BICEP:2021xfz} for $\Lambda$CDM and that in
Ref.\cite{Ye:2022afu} (fullPlanck+BICEP/Keck+BAO+Pantheon) for
AdS-EDE. }
    \label{fig:rns}
\end{figure}

In Fig.\ref{fig:rns}, we plot the $r-n_s$ posterior in
$\Lambda$CDM and AdS-EDE (as an example of EDE) models,
respectively. Following Ref.\cite{Ye:2022afu}, the
AdS-EDE results are obtained with the fullPlanck+BK18+BAO+Pantheon
dataset using the modified versions\footnote{The corresponding
cosmological code is available at:
\url{https://github.com/genye00/class_multiscf}.} of CLASS
cosmology code \cite{Lesgourgues:2011re,Blas:2011rf} and
MontePython-3.4 Monte Carlo Markov Chian (MCMC) sampler
\cite{Audren:2012wb,Brinckmann:2018cvx}. The $\Lambda$CDM results
are directly produced from the public available BK18
chains\footnote{Available at
\url{http://bicepkeck.org/bk18_2021_release.html}.}
(fullPlanck+BK18+BAO) \cite{BICEP:2021xfz}.
%{\color{red} \sout{The $\Lambda$CDM model yields bestfit $\chi_{\rm{Planck}}^2=2778.10$ and $\chi_{\rm{BK18}}^2=537.25$.}} 
AdS-EDE fits the full Planck CMB and
BICEP/Keck18 B-mode data slightly better than $\Lambda$CDM with
bestfit $\chi^2$ improvements $\Delta \chi^2_{\rm{Planck}}=-5.49$
and $\Delta\chi^2_{\rm{BK18}}=-1.32$.

%\begin{table}
%    \begin{tabular}{|c|c|c|}
%        \hline
%        Dataset&$\Lambda$CDM&AdS-EDE\\
%        \hline
%        Planck high-$l$ TTTEEE&2349.64  &2348.16 \\
%        Planck low-$l$ TT&23.61  &20.54 \\
%        Planck low-$l$ EEBB&395.90  &392.74 \\
%        Planck lensing&8.95&11.17\\
%        BK18&537.25  &535.93 \\
%        BAO&5.78  &5.46 \\
%        SNIa&--  &1026.86 \\
%        \hline
%    \end{tabular}
%\caption{Bestfit $\chi^2$ for each likelihood. The $\Lambda$CDM
%bestfit is taken to be the point with the lowest total $\chi^2$
%value in the publicly available chains by BK18.}
%    \label{chi2}
%\end{table}

In Hubble-tension-free AdS-EDE model, $n_s=0.998\pm0.005$, which
requires that $N_*\gtrsim 300\gg \Delta N$. It is interesting to
note that the hybrid uplift to $n_s=1$ generally lowers $r$, which
has the potential to revive inflation models killed by
Planck+BICEP/Keck based on $\Lambda$CDM due to too
large $r$. In Fig.\ref{fig:rns}, when $N_*>300$, we have
$0.993\lesssim n_s\leqslant 1$ and
$r\simeq8/{N_*}\lesssim 0.03$ for chaotic $\phi^2$
inflation, perfectly consistent with current constraints. Since here $r\sim |n_s-1|$ ($r\sim
\left(n_s-1\right)^2$ in Starobinski model), so in certain sense,
$n_s=1$ might also explain the non-detection of $r$ in current
observations.

It is required that the effective field theory responsible for the
evolution of our Universe must be UV-complete, otherwise it
belongs to the swampland. According to (\ref{N}), we have $\Delta
N\sim {\Delta\phi\over \sqrt{2\epsilon}}$, where
$\Delta\phi=\phi_*-\phi_c$. In $\phi^2$ inflation without hybrid
uplift, we have $N_*= \Delta N$ and $\epsilon\sim {1/ N_*}$, so
the field excursion of inflaton is $\Delta \phi\approx \sqrt{2N_*}>1$, contradicting the swampland conjecture $\Delta \phi< 1$. However, for
hybrid $\phi^2$ inflation, we have
\be \Delta \phi\approx \sqrt{2
(\Delta N)^2\over N_*}, \ee
where $\Delta N\approx 60\ll N_*$.
Thus the swampland conjecture $\Delta \phi<1$ requires $N_*\simeq
10^4$.
%\footnote{It might be a concern that $\phi_*\gg 1$ for
%$N_*\gg \Delta N\approx 60$. However, we always can rewrite $\phi$
%as $\phi+\phi_c$ and the effective potential as $V_{\rm{inf}}\sim
%(\phi+\phi_c)^2$, so that initially $\phi_*=\Delta\phi<1$. }
In this case, $r\approx 8/N_*\sim 10^{-3}$, also
consistent with the Lyth bound \cite{Lyth:1996im}.

\textbf{(hybrid) polynomial attractors} In corresponding models,
$V_{\rm{inf}}(\phi)\sim 1-\left({\mu\over \phi}\right)^p$, see recent
Ref.\cite{Kallosh:2022feu}, which was invented in D-brane
inflation \cite{Dvali:1998pa,Burgess:2001fx,Kachru:2003sx}, and
\begin{equation}
n_s-1\approx -\frac{2}{N_*}\left({p+1\over p+2}\right),\qquad
r\approx {8p^2\over \left[p(p+2)N_*\right]^{2p+2\over
p+2}}\mu^{2p\over p+2},
\end{equation}
for $\mu\ll 1$. Thus for e.g. $p=2$, we have $n_s-1=-{3/ 2N_*}$
and $r\simeq N_*^{-3/2}\mu$, which is also
compatible with the $\Lambda$CDM constraint $n_s\approx0.97$ for
$N_*=\Delta N\approx 60$. However, in Hubble-tension-free
cosmologies, $n_s=0.998\pm0.005$, this would require that
$N_*\gtrsim 200\gg \Delta N$ and inflation ends at $N\gtrsim 140$
(60 efolds after the CMB modes exit horizon at $N_*\simeq 200$).
Thus we have $n_s\simeq1-3/2N_*=0.993$, and $r\simeq 5\times10^{-4}\mu$,
which is still consistent with current constraint but even smaller than
that in hybrid Starobinsky model.

\section{Conclusion} \label{sec:conclusion}

The complete resolution of Hubble tension might be
pointing to a scale-invariant Harrison-Zeldovich spectrum of
primordial scalar perturbation, i.e. $n_s=1$ for $H_0\sim
73$km/s/Mpc. We propose a scheme to lift $n_s$ predicted by well-known slow-roll inflation models to $n_s=1$. In
corresponding models satisfying (\ref{ns}), if inflation ends by a
waterfall instability when inflaton is still at a deep slow-roll
region, $n_s$ can be lifted to $n_s= 1$. Particularly, it is found
that chaotic $\phi^2$ inflation ruled out by Planck+BICEP/Keck
\cite{BICEP:2021xfz} can be revived by this hybrid uplift, which
is testable with upcoming CMB B-mode experiments.

The inflation might continue after the waterfall instability. It
is possible that the waterfall instability is caused by the
nucleations and collisions of vacuum bubbles \cite{Linde:1993cn}.
This will yield a sub-horizon stochastic GW background, which
after being reddened by subsequent inflation can
explain the recently observed NANOGrav signal
\cite{Li:2020cjj,Wang:2018caj}. It is also possible that EDE is
the remnant after the waterfall instability along $\sigma$,
so that inflaton, EDE and current dark energy could
live harmoniously together in a landscape. Relevant models
might imprint lots of observable signals to be explored.

Though our discussion is slightly simplified, it highlights a
significant point that until the Hubble tension is solved
completely, it seems \textsf{premature} to claim which model of
inflation is favored or ruled out by current data, since new
physics beyond $\Lambda$CDM might bring unforeseen impact on
primordial Universe.

\paragraph*{Acknowledgments}
This work is supported by NSFC, Nos.12075246, 11690021. The
computations are performed on the TianHe-II supercomputer.

\bibliography{refs}

\end{document}